\documentclass[12pt]{iopart}
\usepackage{graphicx}
\usepackage{hyperref}
\hypersetup{hidelinks}
\usepackage[font=footnotesize]{caption}

\begin{document}

\title[]{STAR-FDTD : Space-time modulated acousto-optic guidestar in disordered media}

\author{Michael Raju$^{1,2,*}$, Baptiste Jayet$^{1}$\& Stefan Andersson-Engels$^{1,2}$}

\address{$^1$ Tyndall National Institute, Lee Maltings Complex, Dyke Parade, Cork, Ireland, T12 R5CP}
\address{$^2$ School of Physics, University College Cork, College Road, Cork, Ireland, T12 K8AF}
\address{$^*$ Author to whom any correspondence should be addressed.}

\ead{\mailto{michaelraju@umail.ucc.ie }}
\vspace{10pt}
\begin{indented}
\item[]November 2023
\end{indented}

\begin{abstract}
We developed a 2D Finite-Difference Time-Domain (FDTD) method for modeling a space-time modulated guidestar targeting wavefront shaping applications in disordered media. Space-time modulation in general (a particular example being the acousto-optic effect) is used here as a guidestar for the transverse confinement of light around the tagged region surrounded by disorder. Together with the guidestar, the iterative optical phase conjugation (IOPC) method is used to overcome the diffusion of light due to multiple scattering. A phase sensitive lock-in detection technique is utilized to estimate the steady-state amplitude and phase of the modulated wavefronts emerging from the guidestar region continuously operating in the Raman-Nath regime. As the IOPC scheme naturally converges to the maximally transmitting eigenchannel profile, one could use the position of the guidestar within the disorder to channelize the maximal transmission through the tagged region. The associated code developed in MATLAB\textsuperscript{\textregistered} is provided as an open source (The MIT License) package. The code package is referred by the acronym STAR-FDTD where STAR stands for \textbf{S}pace-\textbf{T}ime modulated \textbf{A}cousto-optic guidesta\textbf{R}. 
\end{abstract}

%
\vspace{2pc}

%
%
%
\section{Introduction}
Wavefront shaping experiments involving coherent control \cite{RevModPhys.89.015005} of light has been an actively researched area in the past decade. The ability to shape the transport of light through a disordered medium finds its use in a diverse spectrum of applications \cite{gigan2022roadmap} covering photonics in general. Essentially, this includes the focusing of multiple scattered light outside a disorder \cite{vellekoop2007focusing,popoff2011controlling} and also enhancing total transmission through a disorder \cite{popoff2014coherent,hsu2017correlation}. Sometimes it is desirable, especially in the context of biophotonics, to have the focused region to lie within the random medium. Such a geometry will be of interest for tomography applications by creating a small focus of light inside the disordered medium by overcoming the light diffusion due to conventional multiple scattering.  Such a task can be experimentally achieved by incorporating a ``guidestar" \cite{horstmeyer2015guidestar} inside the random medium, which serves as the tagging region to focus upon via wavefront shaping. The guidestar could be based on focused ultrasound \cite{xu2011time,wang2012deep,si2012breaking}, fluorescence \cite{vellekoop2010scattered,vellekoop2012digital,baek2023phase}, optical non-linearity \cite{katz2014noninvasive}, second harmonic radiation from nano-particles \cite{hsieh2010digital}, photo-acoustic guidance \cite{kong2011photoacoustic,chaigne2014controlling,lai2015photoacoustically}, modulated magnetic-particle-guidance \cite{ruan2017focusing}, genetically encoded photochromism \cite{yang2019focusing}, the intrinsic permittivity variations due to a moving target \cite{ma2014time,zhou2014focusing} or the ultrasonic microbubble destruction \cite{ruan2015optical}. The main aim of this paper is to model the effects of using the acousto-optic \cite{korpel1972applied,korpel1996acousto,elson2011ultrasound,wang2004ultrasound,gunther2017review} interaction as the guidestar. The incoherent transport theory aspects \cite{leutz1995ultrasonic,kempe1997acousto,wang2001mechanisms,sakadvzic2006correlation,huang2020investigating}  of acousto-optic modulation involving diffusive multiple scattered light are well studied. On the other hand, coherent wave scattering modeling involving acousto-optic modulation of multiple scattered light, particularly addressing wavefront shaping is a rarely studied area. Such a modeling study involving spatio-temporal modulations of coherent light waves for wavefront shaping applications is presented in this paper.  

Guidestar based on the acousto-optic effect is particularly important for  biophotonic applications including tomographic imaging \cite{wang2012deep,jayet2013optical,liu2015optical,katz2019controlling} and light delivery \cite{ruan2017deep} through real biological tissue. The potential of acousto-optic guidestar to overcome conventional multiple scattering of light to focus through synthetic disordered media has also been demonstrated \cite{xu2011time,si2012breaking,katz2019controlling,si2012fluorescence,judkewitz2013speckle,ruan2020fluorescence}. Here, the acoustic wave propagation undergo very minimal multiple scattering inside tissue like media providing the ability to create an acoustic focus that could be raster-scanned. Acousto-optic interaction occurring mostly in the focal region helps to non-invasively create a label-free tagged region with good depth at a desired location for imaging or light delivery purposes. The multiply scattered light wave  interacting with this acoustic focus (whose size is limited by the ultrasound wavelength) undergoes frequency shift leading to the generation of tagged light only from the acoustic focal region. Thus, the optically-filtered acoustically-tagged light will help in attaining optical imaging contrast which originates due to the biochemical properties of the tagged region. Such an optical contrast can be overlayed on the top of the bio-mechanical imaging contrast obtained with the ultrasound imaging leading to enhancement of conventional ultrasound imaging. The acousto-optic guidestar based wavefront shaping of light further enhances the conventional acousto-optic imaging by overcoming the multiple scattering of light to better focus the light onto the acousto-optic tagged region enhancing light delivery. Such an optimized light focus can also be combined with fluorescence excitation specifically in the tagged region for imaging inside scattering media.

For numerically modelling the effects of the acousto-optic guidestar we employ the Finite Difference Time Domain (FDTD) method. The exact FDTD modeling of the acousto-optic effect is computationally challenging due to the large difference between the temporal frequency of the light and the acoustic waves. This would naturally lead to the need for a large number of time-stepping in the FDTD scheme. Hence the phenomenon Space-Time Modulation (STM) is used in this paper, where the temporal modulation frequency of the guidestar, $\omega_{mod}$, is very high compared to that of the acoustic frequency. As an example, in this paper, $\omega_{mod}$ is a fraction of the light temporal frequency, $\omega_0$, such that $\omega_{mod}/\omega_{0}=0.2$. This provides the computational easiness in reducing the time-steps compared to that of the true acousto-optic interaction. As STM is a more general framework under which the phenomena of acousto-optic effect is a particular example, the basic transport physics remains very similar to each other. The term STM generally refers to the spatio-temporal modulation of the refractive index covering a wide range of temporal and spatial modulation frequencies within the medium, generating the diffraction effects of a traveling refractive index grating. Therefore, the general term STM is preferred over the term acousto-optics, although the main experimental context is the use of acousto-optic effect as a guidestar. Such a STM model implemented in this paper is not just limited to acousto-optic applications, but any form of continuous  spatio-temporal modulations of refractive index localised inside the disordered medium to be used as a guidestar. For a detailed review on various aspects of STM, one may refer \cite{taravati2019generalized,taravati2020space}. 

Previously, a FDTD scheme was reported~\cite{hollmann2013analysis} for modeling an acousto-optic region as a guidestar. Here, a novel perspective of looking at the same problem is presented, which involves a STM region operating in the Raman-Nath regime, the phase sensitive detection of the modulated harmonic wavefronts and iterative optical phase conjugation (IOPC). We made use of the concepts discussed for modeling the STM done by \cite{taravati2019generalized} and adapted it for wavefront shaping application needs as a guidestar in the disordered media. This paper is a standalone document attempting to pedagogically explain the usefulness of the associated code package (STAR-FDTD) mostly from a physics point of view. In addition to this paper, there is a supplementary document acting as a user manual covering various aspects of the code package. 

\section{Theory and methods}
\label{sec:theoryandmethods}
In this paper, light focusing is modeled by the phase conjugation technique through the tagged region. Inspired from various experiments \cite{si2012breaking,katz2019controlling}, the IOPC method is utilized for the enhanced control compared to one-time phase conjugation. The word ``iterative" corresponds to the phase conjugation of waves from the left and the right sides of the disorder, sequentially, starting from an unshaped wave incidence to converge upon the limiting behavior being the spatially confined transport inside the disorder. Such an IOPC method in transmission with an external gain factor, was found to naturally converge upon the maximally transmitting eigenchannel \cite{katz2019controlling} profile through the acousto-optic guidestar. With such kinds of experimental wavefront shaping methods,  \cite{si2012breaking,katz2019controlling} demonstrated the breaking of the acoustic diffraction limit in the conventional acousto-optic imaging highlighting its importance.  

The basic operation of STM guidestar surrounded by disorder is summarized as the following. When a carrier light wave at $\omega_0$, scattered by the disorder interacts with a STM region, temporally harmonic wavefronts emerge due to the interaction. The STM region in the FDTD method is characterized by the space and time-dependent dielectric constant $\epsilon_{r}(r,t)$ which causes the spatio-temporal modulation of the light passing through. Here, $\epsilon_{r}(r,t)$ of the STM region is configured to have only a single temporal modulation frequency, $\omega_{mod}$ (implying continuous operation) and a single spatial modulation frequency $k_{mod}$. The STM region acts as a traveling sinusoidal refractive index grating which temporally modulates the interacting light at a frequency $\omega_{mod}$ and also spatially modulates the light at a frequency $k_{mod}$. The tuning of the parameters $\omega_{mod}$ and $k_{mod}$, together with the functional form of the STM modulation will be discussed later in \sref{sec:BraggVsRaman-Nath}.
Due to the STM modulation, various harmonic wavefront components emerging from the tagged region have temporal frequencies ($..., \omega_{0} -3\omega_{mod}$, $\omega_{0} -2\omega_{mod}$, $\omega_{0} -\omega_{mod}$, $\omega_{0} $, $\omega_{0} +\omega_{mod}$, $\omega_{0} + 2\omega_{mod}$, $\omega_{0} + 3\omega_{mod}$, etc ...) denoted generally as $ \omega_{0} +\textbf{n}\omega_{mod}$, where the bolded \textbf{n} is an integer due to energy conservation in the STM interaction. Estimating the amplitude and phase of one of the modulated harmonic wavefronts (say, at  $\omega_{\textbf{n}=1}=\omega_{0} +\omega_{mod}$) helps to filter out that wavefront component. Such a filtering of the steady-state modulated wavefront at a given $\omega_{\textbf{n}}$ is implemented by a phase-sensitive lock-in detection \cite{nihei2007frequency} scheme to be described later in \sref{sec:PSDforSTM}. The filtering method yields the steady-state amplitude and phase of the modulated wavefront as a function of space in a memory efficient manner. This is followed by the phase conjugation of the modulated wavefront back to the STM region, time reversing the propagation path it took to emerge out. This results in further interaction with the STM region leading to the IOPC scheme to be presented later in the paper.  

\begin{figure}
	\includegraphics[width=\textwidth]{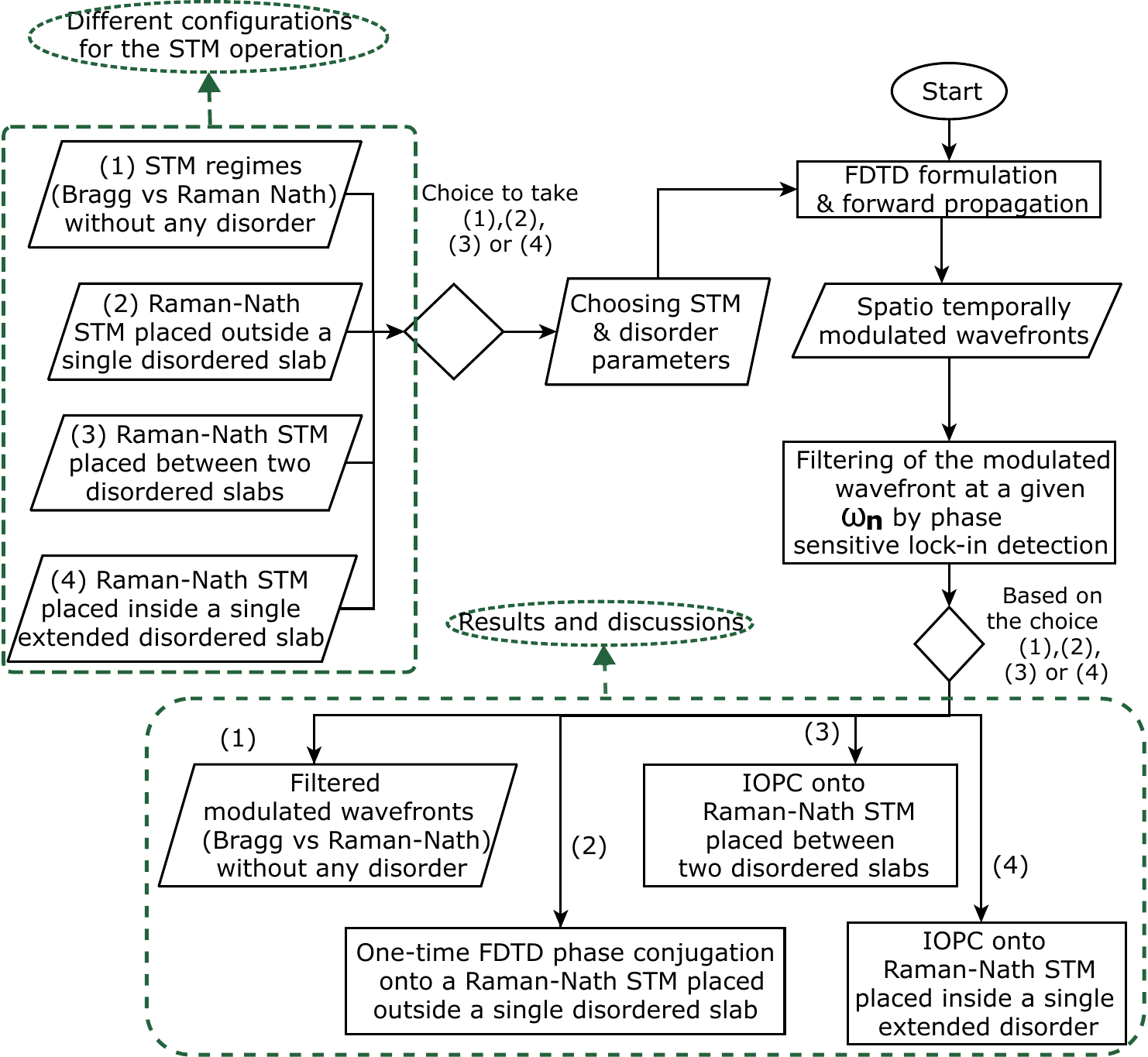}	
	\caption{An overview of the FDTD modeling methods adopted in this paper is shown. The effects of four different STM configurations are studied separately yielding their corresponding results and discussions.}
	\label{fig:overview}
\end{figure}

An overview of the FDTD methods implemented is given in the \fref{fig:overview}. First, the FDTD formulation involving a STM region is described in \sref{sec:FDTDformalism}. Such a FDTD scheme is used for the forward propagation of the light wave involving a STM interaction generating various modulated wavefronts.  The same FDTD scheme is also made use for the phase conjugation of the filtered modulated wavefronts. Filtering of the modulated wavefronts at a given $\omega_{\textbf{n}}$ is described next, by implementing the phase sensitive lock-in detection scheme as given in \sref{sec:PSDforSTM}. Four different STM configurations are studied separately as shown in \fref{fig:overview} yielding their corresponding results and discussions. The first configuration explained in \sref{sec:BraggVsRaman-Nath} is a comparative study between the Bragg and Raman-Nath regimes of STM operations without any disorder involved. Such a study helps to fix the STM parameters to be used later for implementing the guidestar operation in the Raman-Nath regime in presence of disorder.  This is followed by the one-time phase conjugation study where the STM region is placed outside a slab disorder as demonstrated in \sref{sec:onetimePC}. Finally, the IOPC method is explained step by step in \sref{sec:IOPCsteps} where two different disorder geometries are taken. The first disorder geometry for implementing the IOPC method is the scenario where the STM is placed in a scatterer-free region between two disordered slabs.  The second disorder geometry is the case where the STM is placed inside a single extended disordered slab to study the role of multiple scattering. The corresponding results and discussions associated with these four STM configurations are presented in \sref{sec:results&discuss}.

\subsection{FDTD formulation of the wave equation for a Space-Time Modulated (STM) medium}
\label{sec:FDTDformalism}
For the modeling of the STM guidestar, it is essential that the FDTD formulation of the wave equation with time-dependent relative permittivity  $\epsilon_r(r,t)$ is explained. For the computational easiness and simplicity, the 2D scalar wave equation without any loss or gain within the transport medium is considered.
\begin{equation}
	\fl \eqalign{ \nabla^2 E(r,t)&=\frac{1}{c_0^2} \frac{\partial^2}{\partial t^2} \left(\epsilon_r(r,t)E(r,t)\right) \cr
		&=\frac{1}{c_0^2}\left\{\epsilon _r(r,t)\frac{\partial ^2}{\partial ^2t}E(r,t)+ 2 \left(\frac{\partial }{\partial t}\epsilon _r(r,t)\right)\frac{\partial }{\partial t}E(r,t)+E(r,t)\frac{\partial ^2}{\partial ^2t}\epsilon _r(r,t)\right\},         }
	\label{eq:timedomainHelm3}
\end{equation}
where $E(r,t)$ is the real valued wave field and $c_0$ is the free space light wave velocity. Here, the relative permittivity  $\epsilon_r(r,t)$ is time varying in the STM region and time independent in the disorder surrounding it. To obtain a compact form of the discrete version of the above given equation, the following operator notations $D_n[...]$ and $D_n^2[...]$, representing the first order and the second order temporal derivatives  acting on $\epsilon_r(r,t)$, respectively, are used.
\begin{eqnarray}
	&\epsilon_{r_{i,j}}^n=\epsilon_r(z,y,t)|_{t=(n-1)dt},\\
	&D_n\left[\epsilon_{r_{i,j}}^n\right]=\frac{\partial}{\partial t}\epsilon_r(z,y,t)|_{t=(n-1)dt}=
	\left(\frac{\epsilon_{i,j}^{n+1}-\epsilon_{i,j}^{n-1}}{2\Delta_t}\right),\\
	&D_n^2\left[\epsilon_{r_{i,j}}^n\right]=\frac{\partial ^2}{\partial ^2t}\epsilon _r(z,y,t)|_{t=(n-1)dt}=
	\left(\frac{\epsilon_{i,j}^{n+1} -2 \epsilon_{i,j}^n+\epsilon_{i,j}^{n-1}}{\Delta_t^2}\right).
\end{eqnarray}

On a 2D discrete grid, $z$ indexed by $i$ such that $z=(i-1)\Delta_z$, $y$ indexed by $j$ such that $y=(j-1)\Delta_y$, and the time $t$ is indexed by $n$, such that $t=(n-1)\Delta_t$ where $i,j,n \in \left(1,2,3,4, ...\right)$. Discretizing the spatial and temporal derivatives involving $E(r,t)$,  equation \eref{eq:timedomainHelm3} can be approximated as
\begin{equation}
	\eqalign{	\fl \left\{\frac{E_{i,j-1}^n+E_{i,j+1}^n-2 E_{i,j}^n}{\Delta_y^2}\right\}+\left\{\frac{E_{i-1,j}^n+E_{i+1,j}^n-2 E_{i,j}^n}{\Delta_z^2}\right\}=\cr
		\fl  \left.\left.\frac{1}{c_0^2}\left\{\frac{\left(E_{i,j}^{n-1}+E_{i,j}^{n+1}-2 E_{i,j}^n\right) \epsilon _{r_{i,j}}^n}{\Delta _t^2}+ \frac{2 \left(E_{i,j}^{n+1}-E_{i,j}^{n-1}\right) D_n\left[\epsilon _{r_{i,j}}^n\right]}{2 \Delta _t}+E_{i,j}^n D_n^2\left[\epsilon _{r_{i,j}}^n\right]\right.\right.\right\}. }
\end{equation} 
For simplicity, let $\Delta_z=\Delta_y=\Delta_s$ and rearranging the above equation to define parameters $w_0$, $w_1$ and $w_2$ for compactness, 
\begin{eqnarray}
	& \fl	\left\{E_{i-1,j}^n+E_{i+1,j}^n-2 E_{i,j}^n\right\}+\left\{E_{i,j-1}^n+E_{i,j+1}^n-2 E_{i,j}^n\right\}=\nonumber\\
	& \fl	\underbrace{\left(\frac{\epsilon_{r_{i,j}}^n\Delta_s^2}{\left(c_0 \Delta _t\right)^2}\right)}_{w_0} \left(E_{i,j}^{n-1}+E_{i,j}^{n+1}-2 E_{i,j}^n\right) + \underbrace{\left(\frac{\Delta_s^2 D_n\left[\epsilon _{r_{i,j}}^n\right]}{c_0^2 \Delta _t}\right)}_{w_1} \left(E_{i,j}^{n+1}-E_{i,j}^{n-1}\right) + \underbrace{\left(\frac{\Delta_s^2 D_n^2\left[\epsilon_{r_{i,j}}^n\right]}{c_0^2} \right)}_{w_2} E_{i,j}^n .\nonumber\\
	\label{eq:timedomainHelm4}
\end{eqnarray}

Using the defined terms $w_0$, $w_1$ and $w_2$, a time-stepping equation is formed for $E_{i,j}^{n+1}$, in terms of $E_{i,j}^{n}$ and $E_{i,j}^{n-1}$, by rearranging equation \eref{eq:timedomainHelm4} as
\begin{equation}
	\eqalign{	\fl	E_{i,j}^{n+1}=\left(\frac{2 w_0-w_2}{w_0+w_1}\right) E_{i,j}^n-\left(\frac{w_0-w_1}{w_0+w_1}\right) E_{i,j}^{n-1} + \cr
		\left(\frac{1}{w_0+w_1}\right)\left\{\left(E_{i-1,j}^n+E_{i+1,j}^n-2E_{i,j}^n\right)+\left(E_{i,j-1}^n+E_{i,j+1}^n-2 E_{i,j}^n\right)\right\}.    }
	\label{eq:timesteppingUS}
\end{equation}
This time stepping equation forms the core part of the FDTD method for visualizing and estimating the field propagation. The grid size parameters $\Delta_s$ and $\Delta_t$ are chosen to obey the Courant–Friedrichs–Lewy 
stability condition \cite{taflove2005computational} in time-stepping for a given wavelength and minimum refractive index. The space-time modulation resulting in the formation of the guidestar is encoded in the space-time dependence of $\epsilon^n_{r_{i,j}}$. If there is no time dependence for $\epsilon^n_{r_{i,j}}$, then the STM gets nullified and $w_1=w_2=0$.  This is because the temporal derivatives $D_n\left[\epsilon _{r_{i,j}}^n\right]$ and $D^2_n\left[\epsilon _{r_{i,j}}^n\right]$ reduce to zero, resulting in the equation for a temporally stationary medium.

Reflective boundary conditions are implemented on the transverse boundaries while outgoing boundary conditions are used on the transmitting boundaries as explained in the \ref{Appen:ABC}. In addition to these outgoing boundary conditions, there is also a Total-Field/Scattered-Field (TF/SF) boundary condition used as given in \ref{Appen:ABC}. This is for separating out the reflected wave from the total wave due to an incident wave source in the computational domain. 

Next, the wave sources are implemented as a phased array of soft-dipoles for both the excitation of an incident wave or a phase conjugate wave. For the dipole excitation involving point like sources, one point source of the dipole is excited as $E_0(r) cos\left(\omega t+\phi(r)\right)$, while a second point source, placed immediately to the right side, is $E_0(r) cos\left(\omega t + \phi(r) + \pi\right)$, where $E_0(r)$ is the amplitude and $\phi(r)$ the phase of the source. In order to phase conjugate a transmitted wave, we employ time reversal by setting $t=-t$ and time stepping the transmitted wave $ E_{trans}(r,-t)|_{\Gamma_{R}}$ as the phase conjugate wave, injected at the transmission boundary $\Gamma_{R}$. For categorizing the transport regime of the disorder used, the total transmission and reflection of the slab media (without any STM modulation) is estimated by the method given in \ref{Appen:TotalT&R}. This involves the continuity equation for the scalar wave transport and the associated definition of the wave current density $\vec{J}(r,t)$. The temporal and spatial integration of $\vec{J}(r,t)$ for the outgoing waves on the boundary, normalized to the same integration of $\vec{J}(r,t)$  applied to the source wave without the disorder, would yield the total transmission and reflection as given in \ref{Appen:TotalT&R}.     

\subsection{Phase sensitive lock-in detection for estimating the modulated wave fields}
\label{sec:PSDforSTM}
For the phase conjugation of a particular modulated harmonic wavefront, the amplitude and phase of that particular modulated wavefront, at a given $\textbf{n}$ value, have to be determined.  Here, the integer $\textbf{n}$, the temporal diffraction order number, is chosen to be a bold letter $\textbf{n}$ to distinguish from the FDTD time stepping index $n$. Estimating the amplitude and phase of the modulated harmonic wavefront emerging from the STM guidestar for a given $\omega_{\textbf{n}}$ results in the filtering of that modulated wavefront component. For such a filtering, the  method implemented is the phase sensitive lock-in detection scheme reported by \cite{nihei2007frequency} for the wavefront estimation. Such a method is elaborated in \ref{sec:PSDmethod}. As an example, consider the scenario that the transmitted upconverted wavefront at $\omega_{\textbf{n}=+1}=\omega_0+\omega_{mod}$ needs to be filtered leading to the estimation of its amplitude and phase. Then, as per the \ref{sec:PSDmethod}, the temporal cross-correlation terms (given in equation	\eref{correqn:1} and equation \eref{correqn:2}) for $\textbf{n}=1$, taken with respect to the in-phase and quadrature-phase reference waves (given in equation \eref{eq:inphase} and equation \eref{eq:quadphase}), helps in the estimation of the steady-state amplitude and phase of the upconverted wavefront (given by equation \eref{eq:lockinamp} and equation \eref{eq:lockinphase}). Such a lock-in estimation of the steady-state amplitude and phase for the modulated wavefront at a given $\textbf{n}$, is a memory efficient method. Such a method is memory efficient  in comparison to the recording of the time-dependent wave field values for performing time-reversal to focus back at the STM region. The estimated field will be time-reversed later by a dipole-array source placed at the boundary involving the one-time and IOPC phase conjugations. Such a wavefront estimation method is utilized to test the known \cite{taravati2019generalized} diffraction trends in the two regimes of STM operations, namely, the Bragg and the Raman-Nath regimes, reviewed as the following. 

\subsection{Bragg vs Raman-Nath STM diffraction}
\label{sec:BraggVsRaman-Nath}
For modeling the STM without any disorder present in the computational domain, the time-dependent relative dielectric permittivity constant  $\epsilon_r(r,t)$ in the equation \eref{eq:timedomainHelm3}, takes a sinusoidal spatio-temporal modulation \cite{taravati2019generalized} in the STM grating region as
\begin{equation}
	\eta^2(r,t)=\epsilon_{r}(r,t) =\epsilon^{ref}_{r} + \delta \epsilon_{r}\left\{1+sin(k_{mod}y-\omega_{mod}t)\right\},
	\label{eq:STMgrating}
\end{equation}
where $\epsilon^{ref}_{r}$ is the dielectric constant of the reference region surrounding the grating, $\delta \epsilon_{r}$ is the dielectric modulation amplitude, $k_{mod}=2\pi/\Lambda$ is the STM spatial frequency, $\omega_{mod}$ is the STM temporal frequency and $\eta(r,t)$ is the refractive index. Let the average relative dielectric constant of the grating be denoted as $\epsilon^{avg}_r=\epsilon^{ref}_{r} + \delta \epsilon_{r}$. 
The diffraction from a space-time modulated grating is categorized into two main regimes \cite{moharam1978criterion,taravati2019generalized}, namely, the Bragg regime and the Raman-Nath regime.  The Bragg regime is sometimes referred to as the thick-grating regime with a comparatively higher spatial modulation frequency. On the other hand, the Raman-Nath regime is referred to as the thin-grating regime with a comparatively lower spatial modulation frequency. The Bragg regime is characterized by one or two spatial diffraction orders  and the Raman-Nath regime is characterized by multiple spatial diffraction orders.

As there is no sudden step-change cross-over from one regime to another, several works have been published with a goal to define a parameter to distinguish the two regimes of operation. A popular choice is to use the Klein-Cook \cite{klein1967unified} parameter $Q$ to broadly distinguish the regimes. $Q$ is defined as
\begin{eqnarray}
	Q =\left(\frac{k_{mod}}{k_{ref}}\right) \left(k_{mod} L_{mod}\right)=\frac{2 \pi  \lambda  L_{mod}}{\Lambda ^2 n_{ref}},
	\label{eq:KleinCookParameter}
\end{eqnarray}  
where $k_{ref}=(2\pi \eta_{ref})/\lambda$ is the light spatial frequency, $\eta_{ref}$ is the refractive index of the reference region surrounding the STM region and $L_{mod}$ is the longitudinal thickness of the grating.
For the STM modulations presented in here, $\omega_{mod}<\omega_{0}$ and $k_{mod} < k_{ref}$ have to be satisfied. For $\nu=\left(k_{ref}L_{mod}\right) \delta \epsilon_r < 6$ \cite{klein1967unified}, $Q$ can be  used to classify the type of STM regime. For $Q>10$, STM is in the Bragg regime while for $Q$ of the order of unity or below, it can be considered to be in the Raman-Nath regime. Based on the $Q$ parameter, the operating regime of the STM guidestar region is tuned in this paper by selecting appropriate parameters associated with the carrier light wave and the STM region. A detailed modeling study for that is given by Taravati and Eleftheriades \cite{taravati2019generalized}. 

\subsection{One-time phase conjugation onto a Raman-Nath STM guidestar placed outside a disorder} 
\label{sec:onetimePC}
For the one-time phase conjugation from the right side of a disordered slab, the Raman-Nath STM guidestar is placed on the left side of the slab. Here, for the one-time phase conjugation, the Raman-Nath STM has a top-hat spatial modulation envelope. The STM region directly gets excited by the incident beam from the left side. The modulated wavefronts emerging from the STM region gets scattered through the diffusive disorder. Phase sensitive lock-in detection of the transmitted modulated wavefronts is performed. This yields the amplitude and phase of the desired temporal order to be phase conjugated. The same amplitude and phase of the transmitted wave is assigned to an array of time-reversed dipole sources (with the same temporal frequency of the modulated wavefront detected, $\omega_n$) on the right side of the disorder. This is to phase conjugate the wave to be focused on the STM region from where the modulated wavefronts originated.
Also, when the light wave time reverses during the one-time phase conjugation operation, the STM region is switched off. 

\subsection{Iterative optical phase conjugation (IOPC) method}
\label{sec:IOPCsteps}
\begin{figure}[!htp]
	\includegraphics[width=\textwidth]{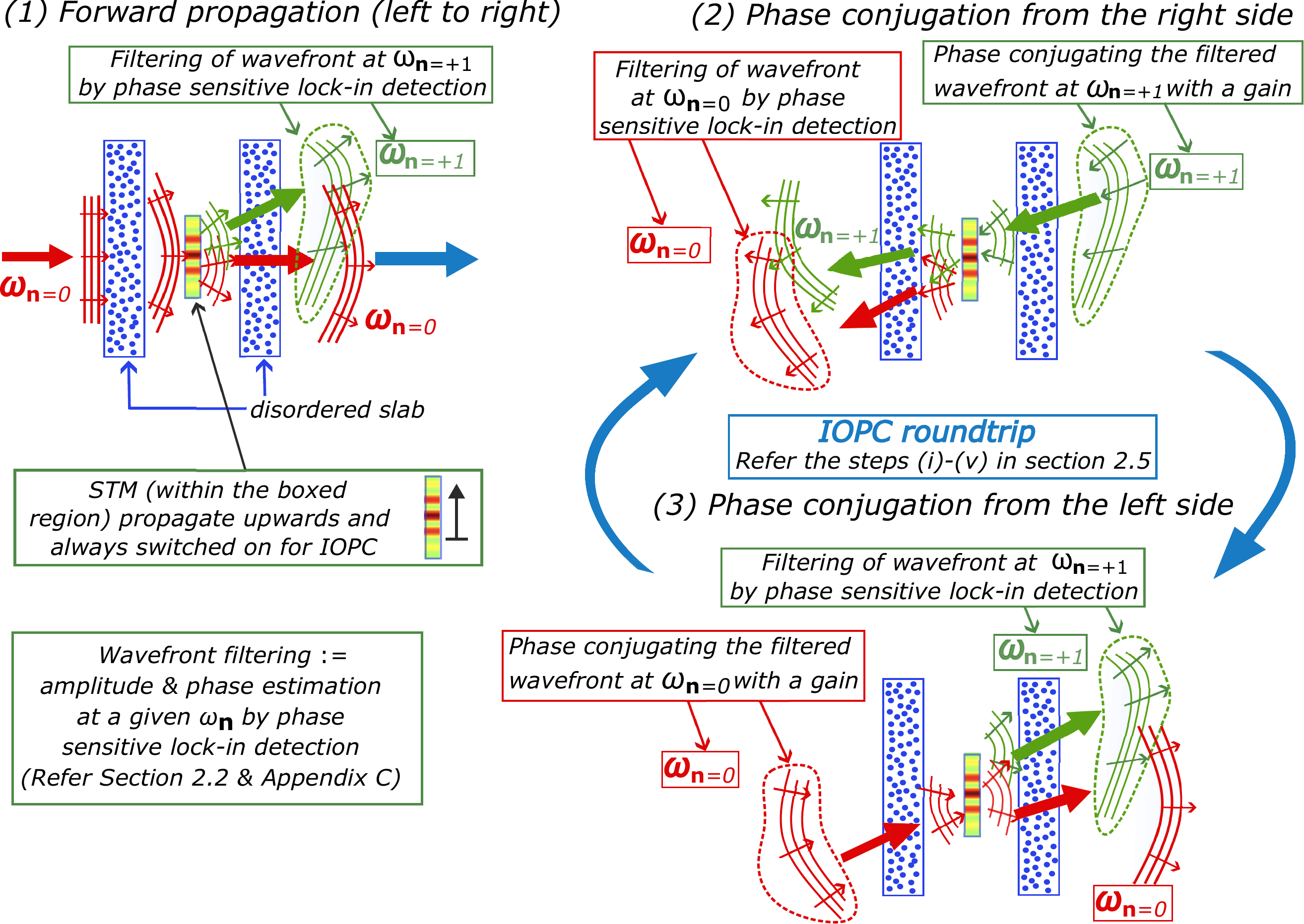}	
	\caption{The main steps involved with the IOPC through a STM region surrounded by disorder are shown. Only those modulated wavefronts with temporal frequencies $\omega_{\textbf{n}=0}$ and $\omega_{\textbf{n}=+1}$ are shown here for simplicity. The detailed breakdown analysis of these steps are given itemized from (i) to (v) in  \sref{sec:IOPCsteps}. \textbf{(1)} Forward propagation from the left to the right side of the disorder is shown where the incident beam at $\omega_{\textbf{n=0}}$ excites the slabs and the STM region. Transmitted upconverted wavefront at $\omega_{\textbf{n=+1}}$ formed due to the STM interaction is filtered. Wavefront filtering, which yields the amplitude and phase of the modulated wavefront at a given $\omega_{\textbf{n}}$, is implemented using the phase sensitive lock-in detection scheme described in \sref{sec:PSDforSTM} and \sref{sec:PSDmethod}.  \textbf{(2)} The filtered upconverted wavefront is phase conjugated from the right side at the same frequency $\omega_{\textbf{n=+1}}$ with a gain factor to focus back at the STM region. The downconverted transmitted wavefront at $\omega_{\textbf{n=0}}$  formed due to the STM interaction is filtered on the left side. \textbf{(3)} The filtered downconverted wavefront is phase conjugated from the left side at the same frequency $\omega_{\textbf{n=0}}$ with a gain factor to focus back at the STM region. The steps (2) and (3) are cyclically repeated for the IOPC process.}
	\label{fig:IPCSTMschematicmargin}
\end{figure} 
The steps taken for the IOPC method involving the STM modulation is summarized in the following steps. The key elements of the steps are shown in the \fref{fig:IPCSTMschematicmargin}. The STM region is switched on all the time during the entire IOPC process. 

\paragraph{Steps taken for the IOPC through a STM region surrounded by disorder}
\begin{enumerate}
	\item Perform a forward propagation from the left to the right side of the disorder at the source temporal frequency $\omega_{\textbf{n}=0}$. The multiply scattered light propagates through the STM region to generate the harmonic wavefronts.
	\item On the right side perform the filtering of the modulated  harmonic wavefront (up-converted wavefront) at the temporal frequency $\omega_{\textbf{n}=+1}=\omega_{0}+\omega_{mod}$. Filtering implies the estimation of the amplitude and phase of the modulated wavefront at a given $\omega_{\textbf{n}}$ value. For the implementation of the filtering, the phase sensitive lock-in detection method described in \sref{sec:PSDforSTM} and \sref{sec:PSDmethod} is used.
	\item The detected wavefront at $\omega_{\textbf{n}=+1}$ is phase conjugated from the right side at the same temporal frequency $\omega_{\textbf{n}=+1}$ with an external gain factor. The external gain factor, which is multiplied onto the phase conjugated wave field amplitude, is set in such a manner that the peak amplitude is always normalised to unity. Such an amplification of the incident phase conjugate wave is performed for every step of IOPC process from both the left and the right side of the disorder. Incorporation of such a gain factor in IOPC is essential for the convergence of the field profile after many iterations, to compensate for the reflective loss during each propagation. Experimentally, such an external gain factor is achieved in the holographic phase conjugation process where an intense reference beam is used to amplify and play-back the weaker recorded wavefront which would be used for phase conjugation. Such a normalised phase conjugate wave from the right, interacts with the STM region and gets up-converted and down-converted in temporal frequency. Only the down-converted wavefront at  $\omega_{\textbf{n}=0}$ by the STM region is now detected on the left side.
	\item The down-converted wavefront $\omega_{\textbf{n}=0}$, emerged only from the STM region and transmitted to the left side of the slab, undergoes phase sensitive detection on the left side at the temporal frequency $\omega_{\textbf{n}=0}$. Such a wavefront is phase conjugated once again with an external gain factor at $\omega_{\textbf{n}=0}$  and set as an incident wave on the left side, completing one round trip.
	\item Several such round trips are repeated phase conjugating the up-converted wavefront ($\omega_{\textbf{n}=1}$) from the right side and the down-converted wavefront ($\omega_{\textbf{n}=0}$) from the left. 
\end{enumerate}

The effects of the disorder geometry on the light focussing during the IOPC wave propagation are studied separately by considering two different disorder configurations. Case 1 uses a disorder geometry where the STM region is sandwiched between two disordered slabs, which is the configuration usually seen in the experiments. The goal of the case 1 geometry is to discuss the IOPC process step-by-step for a given single realization of the disorder. Such a given realization of the disorder has the transmission properties similar to that of a disorder that lies in the diffusion-to-ballistic transition transport regime. Case 2 uses an extended single-slab disorder in the diffusion regime with a STM embedded inside instead of the sandwiched two-slab configuration. This is to study the role of multiple scattering while performing IOPC through a STM guidestar. For both the cases involving IOPC, unlike the one-time phase conjugation scenario presented before, there is a Gaussian spatial modulation along the $Y$ direction on the STM region. This provides a peak value of STM modulation at the center.

\section{Results and discussion}
\label{sec:results&discuss}
\subsection{Testing phase sensitive lock-in detection without disorder : Bragg vs Raman-Nath}
 \begin{figure}
 	\centering
	\includegraphics[width=0.9\textwidth]{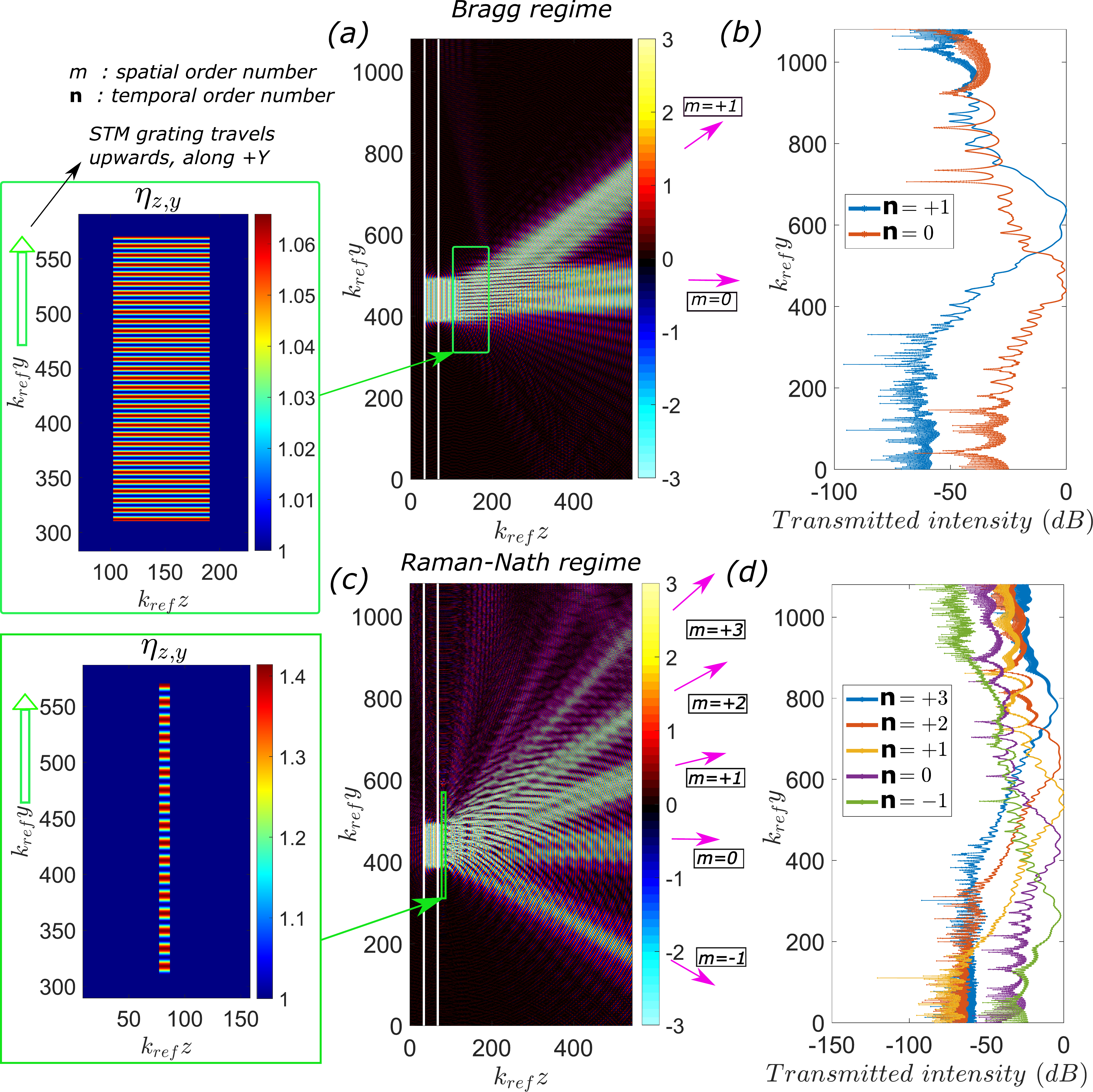}	
	\caption{\textbf{Phase sensitive detection scheme explained in \sref{sec:PSDforSTM} applied onto the known trends \cite{taravati2019generalized} in the spatio-temporal diffracted orders involved with the Bragg and the Raman-Nath regimes, without any disorder present.} \textbf{(a)} The field associated with the wave propagation involving a STM region (enclosed within the green box) operating in the Bragg regime is shown. Two spatial orders indexed by $m=0$ and $m=+1$ are shown. The STM modulation within the green box propagates upward along $+Y$ axis as shown by the direction of the green arrow. For the Bragg regime shown in here, $k_{mod}L_{mod}=64.21$ and $k_{ref}L_{mod}=87.97$, where $L_{mod}$ is the longitudinal thickness (along $z$ axis) of the STM region, $k_{mod}/k_{ref}=0.73$, $\omega_{mod}/\omega_0=0.2$, Klein-Cook parameter $Q=46.88$ and $\nu=k_{ref}L_{mod}\delta \epsilon_r= 5.98$ where $\delta \epsilon_r=0.068$. \textbf{(b)} Phase sensitive detection of the temporally modulated wavefronts with temporal frequencies $\omega_\textbf{n}=\omega_{0}+\textbf{n}\omega_{mod}$ is performed, where $\textbf{n}$ is the temporal order number. Both amplitude and phase of the modulated wavefronts are estimated and only the amplitude is plotted here. \textbf{(c)} The field associated with the Raman-Nath STM diffraction orders are shown. Here, there are multiple spatial diffraction orders indexed by $m$ and the temporal diffraction orders are plotted in \textbf{(d)}. For the Raman-Nath regime shown here, $k_{mod}L_{mod}=3.72$ and $k_{ref}L_{mod}=9.30$, $k_{mod}/k_{ref}=0.4$, $\omega_{mod}/\omega_0=0.2$, Klein-Cook parameter $Q=1.49$ and $\nu=k_{ref}L_{mod}\delta \epsilon_r=4.65$, where $\delta \epsilon_r=0.5$. For both the Bragg and the Raman-Nath regimes, the STM grating takes the form given in equation \eref{eq:STMgrating}. }
	\label{fig:BraggRamanNath}
\end{figure}

Here, the phase sensitive wavefront estimation method is tested for the known \cite{taravati2019generalized} scenarios of STM diffraction without any disorders. The purpose is to verify that the STM grating works as intended, as adding a disorder would scramble and mix various spatio-temporal diffraction orders. Examples of the Bragg and the Raman-Nath regimes, are provided in \fref{fig:BraggRamanNath}(a) and \fref{fig:BraggRamanNath}(c), respectively. The simulation parameters used are given in the figure caption. In \fref{fig:BraggRamanNath}, various temporal diffraction orders are denoted by the index $\textbf{n}$ and the spatial diffraction orders by the index $m$. One may notice that the spatial distribution of the diffraction pattern from the STM region has an asymmetric intensity distribution around the $m=0$ spatial order. This is mainly due to conservation of momentum involving the light wave momentum and the momentum due to the STM modulation which travels upwards along the $+Y$ axis. For further details, refer to \cite{taravati2019generalized}. If the STM region was propagating downwards along $-Y$ direction, then the spatial orders seen before diffract downwards. For detecting the amplitude and the phase of the $\textbf{n}^{th}$ modulated wavefront with the temporal frequency $\omega_{0}+\textbf{n} \omega_{mod}$, a phase sensitive lock-in detection scheme covered in \sref{sec:PSDforSTM} is employed as shown in \fref{fig:BraggRamanNath}(b) and \fref{fig:BraggRamanNath}(d). 

In \fref{fig:BraggRamanNath}, one may notice that for both the Bragg and the Raman-Nath gratings, there is only one single temporal frequency $\omega_{mod}$ at which the STM gratings oscillate. In addition to that, there is only one single spatial frequency $k_{mod}$ at which the STM gratings vary spatially along $Y$. Still, several temporal diffraction wavefront orders ($\textbf{n}=..,-1,0,+1,+2,+3, etc..$) are generated as shown in \fref{fig:BraggRamanNath}(d) for the Raman-Nath regime. As these trends (shown in \fref{fig:BraggRamanNath}) are expected according to \cite{taravati2019generalized}, it can be taken that the FDTD coding for the STM grating is working as intended. 
\subsection{One-time phase conjugation onto a Raman-Nath STM guidestar with disorder}
\begin{figure}
	\centering
	\includegraphics[width=\textwidth]{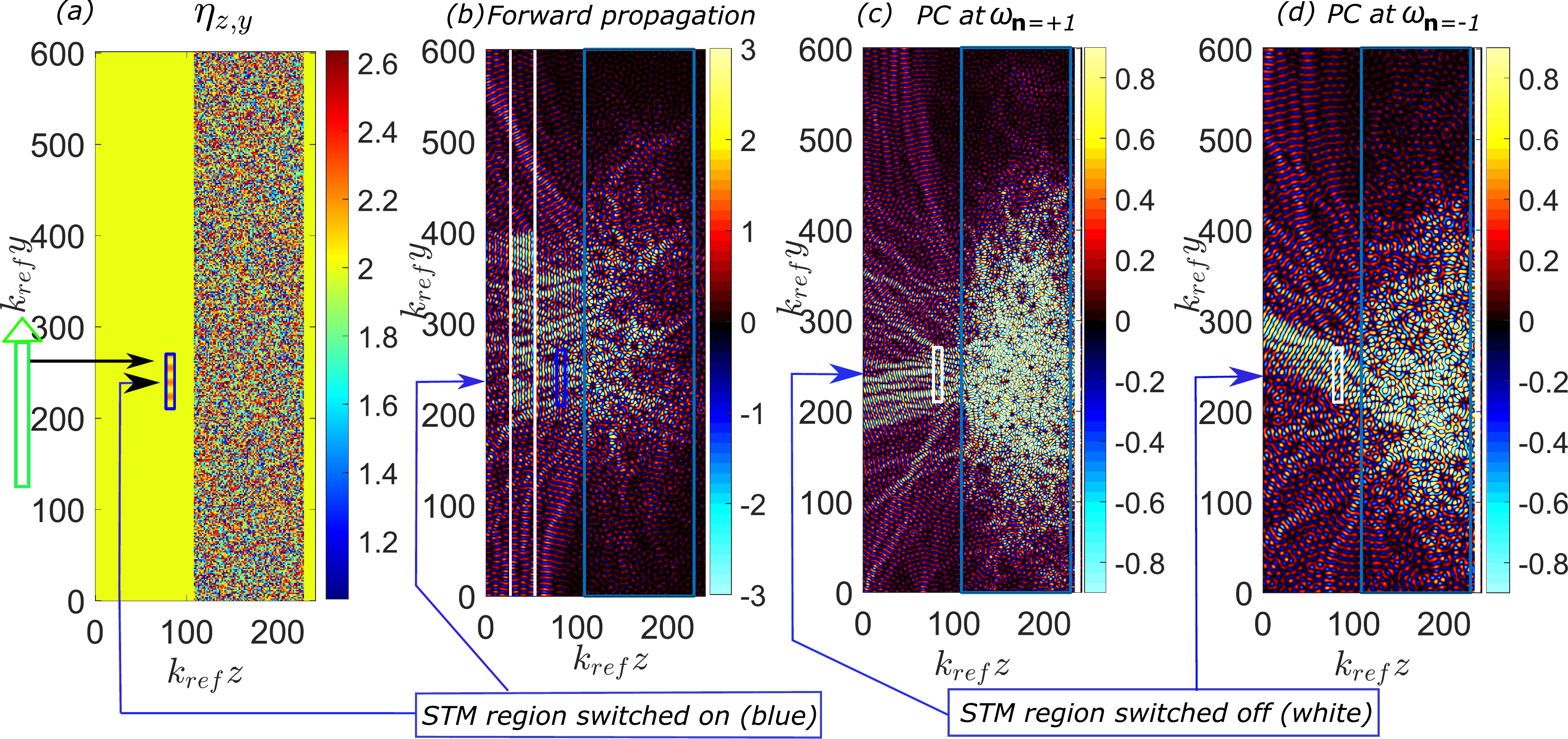}	
	\caption{\textbf{Forward propagation and one-time phase conjugation (PC) involving a STM region operating in the Raman-Nath regime.} \textbf{(a)} Raman-Nath STM travelling upwards (direction indicated by the green arrow) is placed on the left side of a slab disorder. Refractive index $\eta(z,y)$ is plotted which has a top-hat spatial modulation envelope at the STM region along the Y direction. Here, $k_{mod} L_{mod}=3.74$, $k_{ref}L_{mod}=9.35$, $Q=1.5$ and $\omega_{mod}/\omega_{src}=0.2$~\textbf{(b)} Forward propagation at $\omega_0$ is shown, where the incident beam directly excites the STM region (switched on, shown in blue colour, placed at the left side of the slab). Both the incident beam at $\omega_0$ and the modulated wavefronts from the STM region, undergoes multiple scattering due to the disorder on the right. For the disorder without the STM region, the total transmission $T=0.15$, therefore in the diffusion regime. \textbf{(c)} After phase sensitive detection of the $\omega_{\textbf{n}=+1}$ wavefront, it is phase conjugated (only for once) at the same $\omega_{\textbf{n}=+1}$ frequency, to focus back at the STM location which is switched off (shown in white colour). Here,  $T=0.61$. \textbf{(d)} Similarly, the one-time phase conjugation of the $\omega_{\textbf{n}=-1}$ wavefront at the same  $\omega_{\textbf{n}=-1}$ frequency is shown, where $T=0.70$.}
	\label{fig:RamanNathPC}
\end{figure}

As the next step, a disorder is added for using the Raman-Nath STM grating as a guidestar for phase conjugation applications as shown in \fref{fig:RamanNathPC}(a). Using the phase sensitive detection scheme given in \sref{sec:PSDforSTM}, the amplitude and the phase of the $\textbf{n}^{th}$ modulated wavefront with the temporal frequency $\omega_{0}+\textbf{n} \omega_{mod}$ is estimated during the forward propagation as shown in \fref{fig:RamanNathPC}(b). An incident beam larger than the STM region is used intentionally, to verify that the phase conjugated wave due to the modulated wavefronts converge back to the smaller STM region and not to the larger source region.
In \fref{fig:RamanNathPC}, where only the one-time phase conjugation is dealt, the STM region is switched-on (enabling spatio-temporal modulation) only during the forward propagation shown in \fref{fig:RamanNathPC}(b). The enabling of the STM action is indicated by the blue-coloured box enclosing the STM region, yielding the modulated transmitted light. \Fref{fig:RamanNathPC}(c) and \fref{fig:RamanNathPC}(d) demonstrate the focusing through one-time phase conjugation onto the STM guidestar. Both the wavefronts corresponding to $\textbf{n}=+1$ (the upconverted) and $\textbf{n}=-1$ (the downconverted) are phase conjugated separately as shown in the \fref{fig:RamanNathPC}(c) and \fref{fig:RamanNathPC}(d), respectively. While performing these one-time phase conjugations from the right side of the disorder, there is no spatio-temporal modulation at the STM region on the left side of the disorder. The lack of the spatio temporal modulation is shown as a white box, which is kept as a position marker where the STM action existed previously during the forward propagation. The phase conjugation of the wavefronts corresponding to $\textbf{n}=+1$ (the upconverted) and $\textbf{n}=-1$ (the downconverted) are performed to demonstrate the directionality of the phase conjugation involved with respect to the differently modulated wavefronts referred by their temporal modulation order. In regards to the \fref{fig:BraggRamanNath}, one could observe that the $\textbf{n}=+1$ temporal order diffracts upwards during the forward propagation with respect to the $\textbf{n}=0$ temporal order, as the STM grating is traveling upwards along $+Y$. Therefore, one should expect that when the same temporal order $\textbf{n}=+1$ is phase conjugated, it gets time reversed along the same direction from which the diffraction order originally emerged. On the other hand, for the  $\textbf{n}=-1$ temporal order, it diffracted downwards from the STM region with respect to the $\textbf{n}=0$ order on the forward propagation. Hence, for the phase conjugation of $\textbf{n}=-1$ temporal order, it time-reverses in the same direction from which the modulated wavefront originated. This is true, even in presence of disorder as seen in \fref{fig:RamanNathPC}(c) and \fref{fig:RamanNathPC}(d). 
\begin{figure}
	\includegraphics[width=\textwidth]{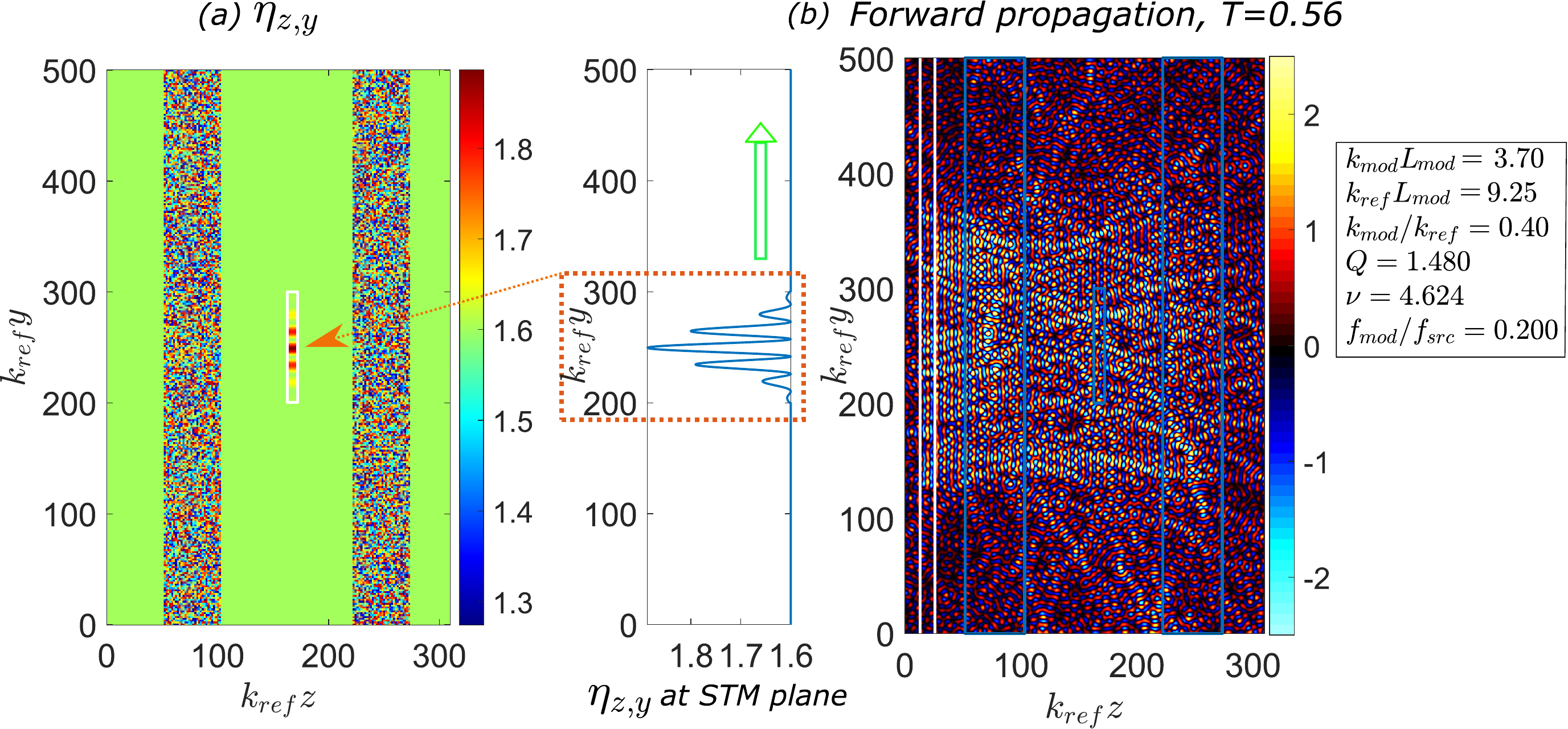}
	\caption{\textbf{Geometry of the Raman-Nath STM modulation in presence of two slab disorders for the IOPC process is shown.} \textbf{(a)} The refractive index associated with the computational domain is shown, where a Gaussian spatial modulation travelling upwards along the $+Y$ axis is placed between two slab disorders. \textbf{(b)} The forward propagation from the left side of the disorder when the STM region is switched-on is plotted. The total transmission value $T$ given is for the combination of the slab disorders when the STM region is switched off.}
	\label{fig:geometryIOPCSTMTimeDom}
\end{figure}
\subsection{Iterative phase conjugation method for the STM region placed between two disordered slabs}
\label{sec:IOPCsandwich}

\begin{figure}
	\includegraphics[width=\textwidth]{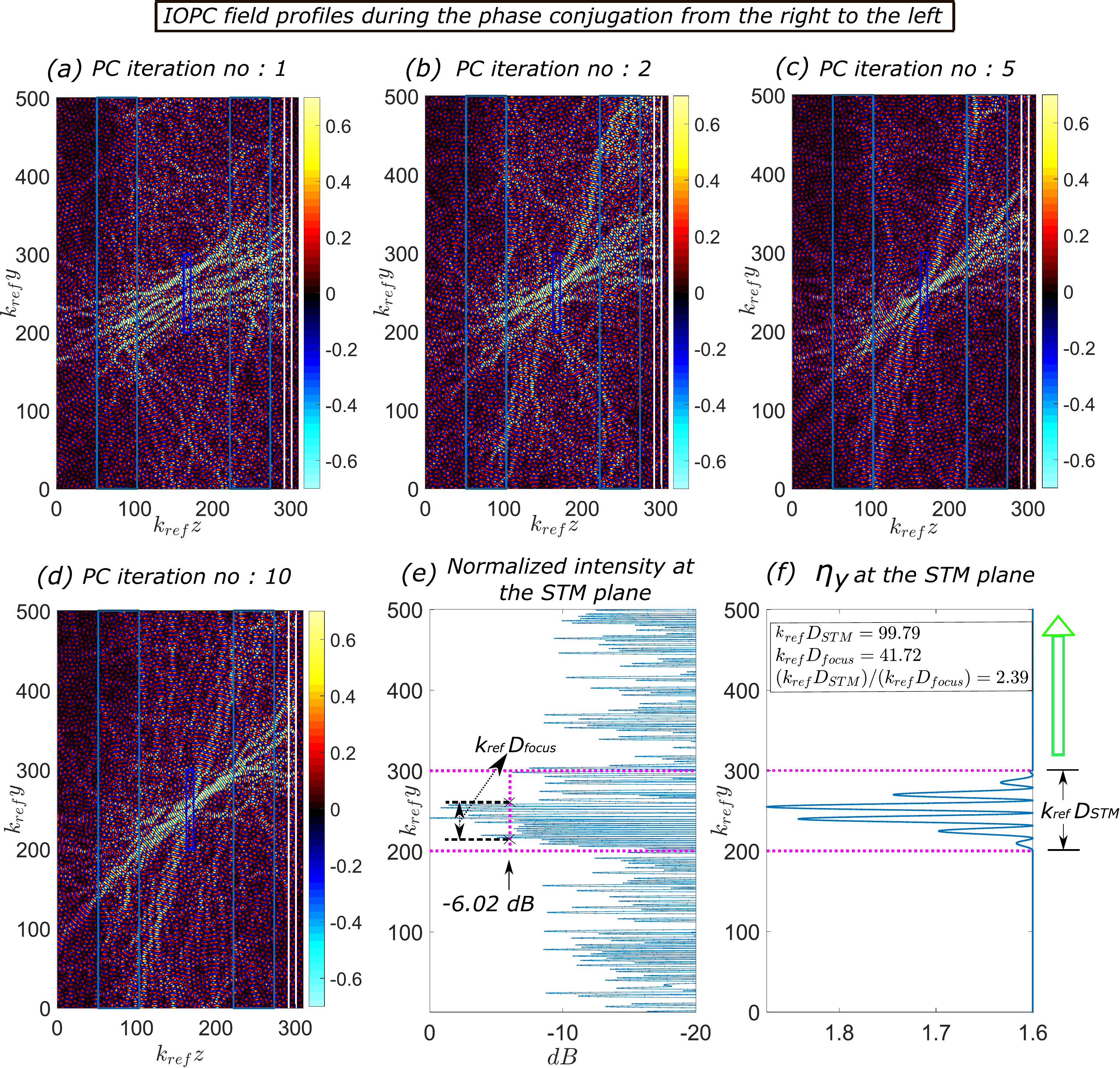}
	\caption{\textbf{Field profiles after several round trips of IOPC through a STM region placed between two disordered slabs.} \textbf{(a)} First phase conjugation of the up-converted wavefront from the right side at $\omega_{\textbf{n}=+1}$ is shown. Subsequent field profiles due to phase conjugation from the right side at $\omega_{\textbf{n}=+1}$ after \textbf{(b)} 2 roundtrips, \textbf{(c)} 5 roundtrips and \textbf{(d)} 10 roundtrips.  
		\textbf{(e)} The intensity of the focussing due to the phase conjugate wave from the right during the $10^{th}$ iteration is shown. The intensity is normalised with respect to the peak intensity. $k_{ref} D_{focus}=41.72$ corresponds to the $-6.02~dB$ point on both sides of the focal peak. \textbf{(f)} The transverse size of the STM modulation is shown where $k_{ref}D_{STM}=99.79$. Therefore $D_{STM}/D_{focus}=2.39$. These plots numerically demonstrates the idea that IOPC has the potential for breaking the acoustic diffraction limit in the conventional acousto-optic imaging. This is in agreement with the experimental observations given by  Katz et al \cite{katz2019controlling} who used acousto-optic eigenchannels for focussing the maximally transmitted eigenchannel and the Ke Si et al \cite{si2012breaking} who used iterative phase conjugation. Also, all the field plots have clipped colour map limits, to make the focussing transport paths and the associated background visible with sufficient contrast. }
	\label{fig:STMPCR2L&L2R3&5&10}
\end{figure}
In this section, the IOPC implementation of the STM guidestar, surrounded by two slab disorders is discussed. For the IOPC process, a Gaussian modulation (shown in \fref{fig:geometryIOPCSTMTimeDom}(a)) is added on the top of the STM profile so that the modulation peaks at the center region. The combined total transmission of the two disordered slabs used for the IOPC process is estimated by the forward propagation shown in \fref{fig:geometryIOPCSTMTimeDom}(b).  For the IOPC process, the STM action is always switched on (indicated by the blue coloured STM enclosure in \fref{fig:STMPCR2L&L2R3&5&10}), because IOPC method needs the phase conjugate wave to be modulated in every pass of the round trip. Upon the progress of round trips where the up-converted wavefront is phase conjugated from the right and the down-converted wavefront phase conjugated from the left, the focus becomes tighter along the transverse dimension. The progression of the IOPC process is shown in \fref{fig:STMPCR2L&L2R3&5&10}(a)-(d) during the 1$^{st}$, 2$^{nd}$, 5$^{th}$ and 10$^{th}$ iterations, respectively. The focus becomes even narrower than the transverse width of the Gaussian modulated STM region as shown in \fref{fig:STMPCR2L&L2R3&5&10}(e) and \fref{fig:STMPCR2L&L2R3&5&10}(f). Such a phenomena was experimentally demonstrated by \cite{si2012breaking,katz2019controlling} involving IOPC and acousto-optic eigenchannels, respectively. In the conventional acousto-optic imaging without any wavefront shaping involved, the resolution is theoretically limited by the diffraction limit of the acoustic wave (corresponding to the entire STM region in this paper). In presence of wavefront shaping, especially with IOPC or eigenchannels, such a conventional acoustic diffraction limit can be broken for the acousto-optic imaging as demonstrated in the \fref{fig:STMPCR2L&L2R3&5&10}(e). 

The IOPC method in this section required the Gaussian spatial-modulation-envelope along the $Y$ direction (shown in 	\fref{fig:geometryIOPCSTMTimeDom}(a)) to iteratively reduce the width of the focus at the STM region. This is because, due to the Gaussian nature of the STM modulation, most of the modulation occurs around the peak. With each roundtrip of the IOPC, the light paths that pass through the STM region that increasingly generates the modulated light transmission would preferentially survive the roundtrip filteration for the tagged light. Such a preferential selection would cause the IOPC propagation to focus most of its power through the STM peak modulation region to maximise the tagged light transmission. Such a role of the Gaussian shaped acousto-optic guidestar has been experimentally demonstrated by \cite{katz2019controlling} in which the acousto-optic transmission eigenchannels with transmissions in the decreasing order were excited. It was observed that the maximally transmitting acousto-optic eigenchannel tightly focussed in the STM peak modulated region. As IOPC through a STM region would converge \cite{katz2019controlling} to the transmission profile of the maximally transmitted STM eigenchannel, a tight focus is attained in this paper with the IOPC method. 

For the one-time phase conjugation study shown in \fref{fig:RamanNathPC}, a top-hat shaped modulation envelop was chosen. It is because in that case, the purpose was not to show the iterative tightening of the phase conjugated wave in the STM region. Instead, the goal was to show that only the modulated plane wavefront, which emerged only from the STM region, was phase conjugated back at the same region from which it originated. The top-hat profile of the STM modulation would generate plane-wave-like phase conjugate wave, which extends only upto the width of the STM region as seen in the  \fref{fig:RamanNathPC}(c) and \fref{fig:RamanNathPC}(d).  With respect to the attainment of a tighter focus in the STM region, the IOPC scheme with a Gaussian spatial modulation is the preferred method compared to that of the one-time phase conjugation method with a top-hat or a Gaussian spatial modulation. It can be observed that for the one-time phase conjugation onto the STM region, whether it be based on the top-hat spatial modulation given in \fref{fig:RamanNathPC}(c) and \fref{fig:RamanNathPC}(d) or whether it be due to a Gaussian spatial modulation shown in \fref{fig:STMPCR2L&L2R3&5&10}(a), the focussed wave is wider compared to that of the IOPC focal profile shown in \fref{fig:STMPCR2L&L2R3&5&10}(d) and \fref{fig:STMPCR2L&L2R3&5&10}(e) where it is tightly focussed.

It needs to be emphasized that the results shown in \fref{fig:STMPCR2L&L2R3&5&10} are obtained for the two-slab geometry (given in \fref{fig:geometryIOPCSTMTimeDom}(a)) with a combined net transmission of 0.56, where the STM region was placed in the disorder-less region between the slabs. A disorder with such a transmission could be catergorized as being in the diffusion-to-ballistic transition transport regime. This would naturally lead to the question of how a single extended diffusive slab with a STM region embedded inside behaves under IOPC and how the degree of multiple scattering plays a role in the focussing. Such a topic is discussed as the following.

\subsection{IOPC for a STM inside a single extended disorder : The effect of multiple scattering}
\label{sec:IOPCsingledisorder}
\begin{figure}
	\includegraphics[width=\textwidth]{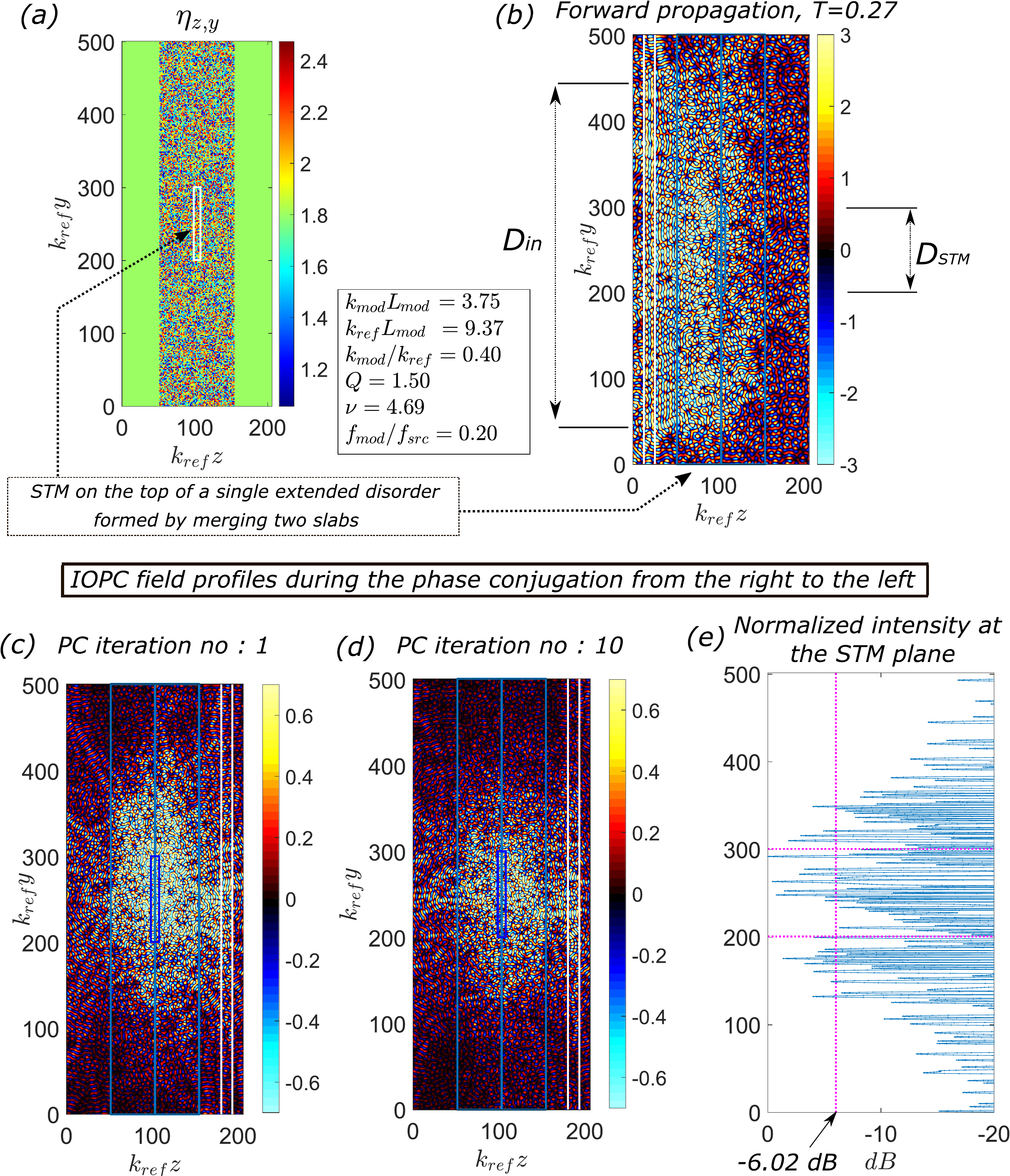}
	\caption{\textbf{Field profiles after several round trips of IOPC through a STM region placed inside a diffusive slab disorder}. \textbf{(a)} The two slab geometry taken before is merged here to create a single extended disorder by setting the separation between the two slabs to be zero. Position of the Raman-Nath STM is enclosed in the white box. \textbf{(b)} Forward propagation due to the incident beam wave starting from the left is shown. The diffusive slab has an unshaped transmission of 0.27 and the beam width $D_{in}$ is taken relatively larger than the STM transverse width $D_{STM}$. \textbf{(c)} Phase conjugation (PC) from the right side during the first iteration and the \textbf{(d)} $10^{th}$ iteration are shown. \textbf{(e)} The normalized intensity at the STM plane after the $10^{th}$ iteration is shown. The transverse width of the STM region and the -6.02 dB intensity value are marked.}
	\label{fig:CombinedPCR2L}
\end{figure}

Here, instead of the two-slab disordered geometry taken before, a single extended slab disorder is considered as shown in \fref{fig:CombinedPCR2L}(a). The STM acts on the top of the disorder refractive index inside the slab. The disordered slab has the unshaped transmission properties corresponding to that of the diffusion regime. Here, the STM parameters taken are similar to the values taken before for the two-slab geometry.
IOPC is performed on the slab disorder with STM region always switched-on, following the same method as discussed before (\sref{sec:IOPCsteps}). IOPC process starts by a plane beam incidence from the left side of the disorder as shown in \fref{fig:CombinedPCR2L}(b). The field profiles inside the slabs due to the phase conjugation from the right to the left side of the slab, in the first and the $10^{th}$ round-trip of the IOPC process, are plotted in \fref{fig:CombinedPCR2L}(c) and \fref{fig:CombinedPCR2L}(d), respectively.  The peak-normalized intensity inside the disorder at the STM region during the iteration number 10 is given in \fref{fig:CombinedPCR2L}(e). This is to compare the peak-to-background contrast with that of two-slab scenario given in \fref{fig:STMPCR2L&L2R3&5&10}(e).  

It can then be observed that the transverse width of the focal region for the single diffusive slab (\fref{fig:CombinedPCR2L}(e)) got increased compared to that of the two slab scenario (\fref{fig:STMPCR2L&L2R3&5&10}(e)). This could be due to the increase in the degree of multiple scattering around the STM region. With the STM region embedded inside a single diffusive slab, the multiple scattering paths cross each other and get spread in-and-around the STM region in the extended slab. The phase conjugate wave try to retrace the same multiple scattering paths as in time-reversal. For the two-slab geometry considered in \fref{fig:STMPCR2L&L2R3&5&10}, the STM region existed in the scattering-free region held between the two slabs, preventing the buildup of dense multiply-scattered paths within the STM region. Numerical discretization errors associated with the simple FDTD method presented here, together with the lock-in detection of the comparatively weaker STM modulated signal from a diffusive slab, may also contribute to the diffusive spreading of time-reversed light within the focal region. Further investigation in the future is needed to cross-verify whether there exist a fundamental limit based on the transport physics of the problem (independent of the numerical method used) for the focussing contrast attained with respect to the degree of multiple scattering. 

On a positive note, it can also be observed from \fref{fig:CombinedPCR2L}(d) and \fref{fig:CombinedPCR2L}(e) that even in presence of multiple scattering, still there is the confinement of light along the transverse dimension (along $Y$) inside the slab region. Hence, even though the focussing contrast decreases around the STM region for the diffusive disorder, the phase conjugate wave after several rounds of IOPC, still remains confined inside the multiply scattering diffusive disorder along the transverse dimension. The IOPC process started with a relatively wide plane beam incidence as shown in \fref{fig:CombinedPCR2L}(b) and the outgoing modulated waves were fully phase conjugated from both sides without any loss associated with the wave collection aperture. This still gave the transverse confinement of waves inside the disorder around the STM region. Therefore, even though the loss of contrast due to increased multiple scattering during IOPC could be considered as a negative result, the control of transverse confinement of light in the diffusive sample using STM seems to be a positive result. 

\section{Conclusion}
The paper implemented the IOPC method to model the confined transmission through the STM guidestar which is equivalent to the 
maximally transmitting eigenchannel through the tagged region.
Such an indirect method of converging onto the eigenchannel profile via IOPC was done mainly because the frequency domain modeling of a transmission matrix incorporating a STM was a challenging problem.  Numerical estimation of such a STM transmission matrix and the description of the properties of eigenchannels are yet to appear in the literature as per the best of our knowledge. Modeling the transmission matrix with the STM in the time domain was also  computationally tedious due to the large number of individual FDTD runs needed for many incident modes, achieving steady-state. Thus the IOPC scheme was felt by us to be a computationally reasonable method with the available resources to converge upon the confined transmission through the STM region. 

The presented FDTD method may be of interest to wavefront shaping researchers to be used as such or to be modified for other applications as FDTD techniques are quite easy to be tailored for custom applications. The presented code package could also be used by researchers new to the field of wavefront shaping to explore and visualize various aspects of wavefront shaping (with or without STM) as a learning tool. Also, the authors believe that the presented methods could be considered as a motivation for other modeling researchers to investigate on numerically efficient schemes to estimate the STM transmission matrix and the associated eigenchannels. Such a maximum transmitting STM eigenchannel may provide signatures of transverse localization. Although it is demonstrated using the modeling that the IOPC method has the capacity to beat the acoustic diffraction limit in certain  scattering regimes, the role played by the degree of multiple scattering also needs to be taken into account while estimating focusing contrast within the diffusive disorder. 

We would also like to point out that in this paper the flux conservation is accounted only when the STM region is switched off. The flux conservation with respect to the total transmission and reflection was valid only in the cases for estimating the transmission profiles of slabs and also while estimating transmissions during one-time phase conjugations. Further study is needed to account for the energy conservation while the STM modulation is switched on. This is because, it seems that the STM modulation performs work and adds energy to the light scattering during the STM interaction process. May be a generalized continuity equation needs to be incorporated to address the work done by the STM region on the light wave. This is to address the total energy budgeting of the STM modulated system which needs to be explored in the future. That is why, this paper mostly focused on estimating only the contrast of focusing, normalized to the peak intensity. 

\section*{Code repository and supplementary document}
The open source (The MIT License) code  packages  \cite{Raju_Matlab_code_packages_2024} along with the supplementary document is hosted in Github (\url{https://github.com/michaelraju/STAR-FDTD.git}) and also in Zenodo (\url{https://doi.org/10.5281/zenodo.10969149}). 

\section*{Author contributions}
M.R. developed the numerical methods and the associated code packages. B.J. and S.A.E. supervised M.R's PhD project. All authors contributed in the discussion, validation and interpretation of the results and the methods. M.R wrote the manuscript with the support from other authors.

\ack
Science Foundation Ireland (SFI) funded the research through S.A.E's professorship grant “Novel applications and techniques for in-vivo optical imaging and spectroscopy” (SFI/15/RP/2828 and SFI/22/RP-2TF/10293). Our special thanks to Prof. Sylvain Gigan (Sorbonne Universit\'{e}) for the stimulating discussions we had regarding the topic covered in this paper. Also, we thank the reviewers and the guest editors for their valuable comments and suggestions. 

\appendix
\section{Derivation for the FDTD Absorbing Boundary Condition (analytical) at $z=0$ and $z=z_{end}$}
\label{Appen:ABC}

Three kinds of boundary conditions are employed. The first kind is the reflecting (mirror) boundary condition which could be easily implemented in the FDTD routine by setting the $E^{n}_{i,j}=0$ at the desired boundary position at all time values. Here, the transverse boundary condition at $y=0$ and $y=W$ is set to be the reflecting boundary condition. The second kind of boundary condition implemented is at the longitudinal boundary planes at $z=0$ and $z=z_{end}$, to cause the scattered waves to be outgoing waves. The generalization to the second order Mur Absorbing Boundary Condition ($ABC$) (Refer section 6.3.3, ``Trefethen-Halpern Generalized and Higher-Order $ABC$s" of \cite{taflove2005computational} for details) are used.

\begin{figure}
	\includegraphics[width=\textwidth]{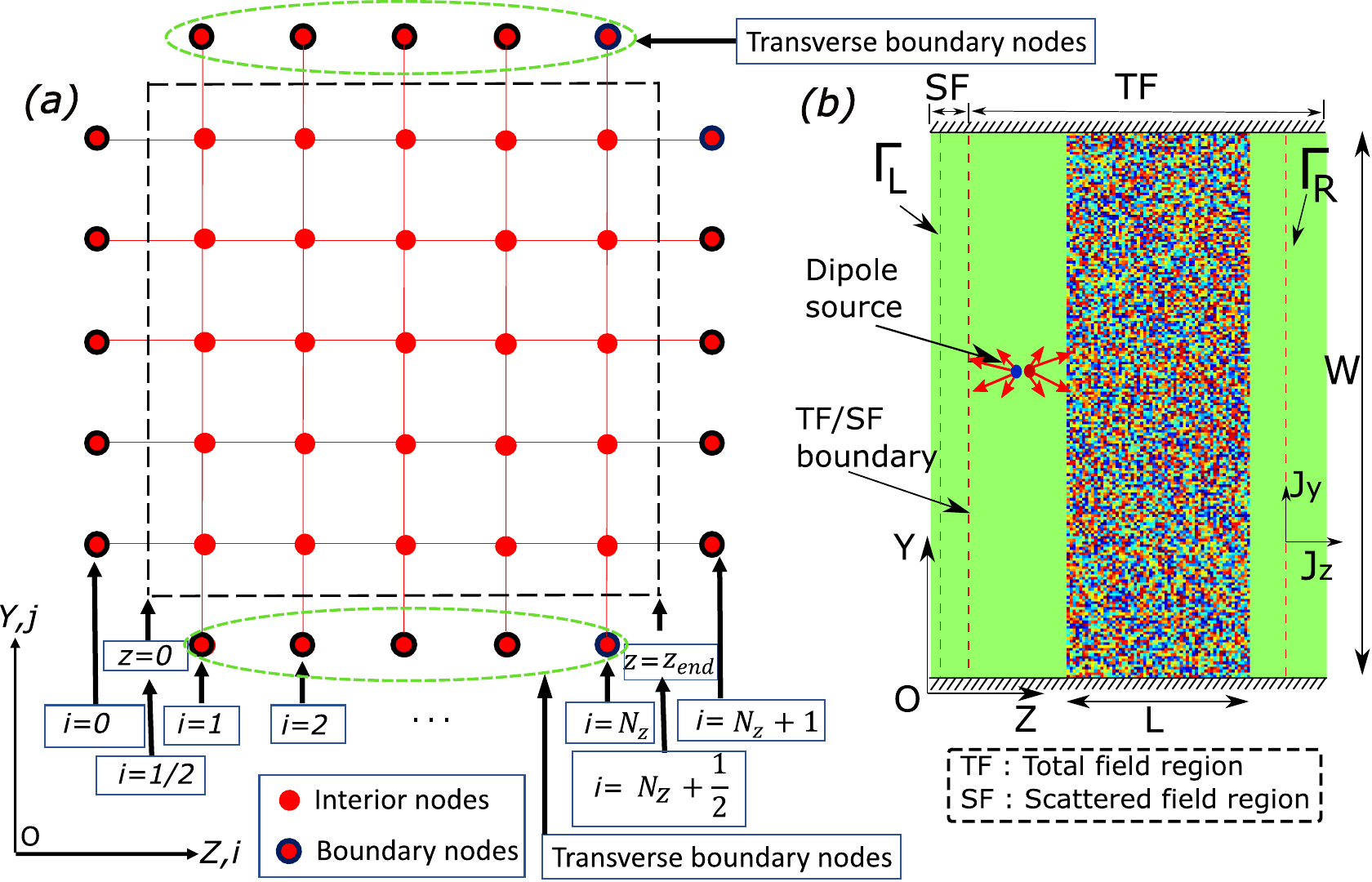}
	\caption{\textbf{Schematics for explaining the boundary conditions.} \textbf{(a)} The grid system and the associated nodes used for the FDTD formulation are shown. $E$ field values on the transverse boundary nodes are forced to be zero for implementing the reflecting boundary condition. On the boundary nodes other than the transverse nodes, outgoing ABC boundary conditions are implemented as per \sref{Appen:ABC}. \textbf{(b)} For a source placed on the left side of the slab medium, the TF/SF areas are marked. SF region contains only the reflected light from the slab as the source field is subtracted from the total field. In this case, the transmitted and reflected wave current component $J_z$ are estimated at the right and the left boundaries $\Gamma_{R}$ and $\Gamma_{L}$.}
	\label{fig:gridsystemABC}
\end{figure}

ABC is incorporated at $z=0$ and $z=z_{end}$ as shown in the  \fref{fig:gridsystemABC}(a), at half-way between interior nodes and boundary nodes. Hence, special care has to be taken to define derivates at $i=\frac{1}{2}$ and $i=N_z + \frac{1}{2}$. (All other definitions of derivative operators were evaluated at the interior grid nodes and not at half-distances from the grid nodes). The goal is to obtain a time-stepping equation for the boundary node fields $E_{0,j}^{n+1}$ and $E_{N_z+1,j}^{n+1}$.
\subsection{FDTD formulation of the ABC at $z=0$}
\begin{equation}
	\frac{\partial ^2E(r,t)}{\partial z\partial t}-\frac{p_0}{c_{ref}}\frac{\partial ^2E(r,t)}{\partial t^2}+p_2c_{ref}\frac{\partial ^2E(r,t)}{\partial y^2}=0.
\end{equation}
Upon discretizing the previous equation and rearranging it to obtain $E_{0,j}^{n+1}$
\begin{equation}
	\eqalign{\fl	E_{0,j}^{n+1}=-E_{1,j}^{n-1}+\frac{c_{{ref}}\Delta_t-p_0 \Delta_z}{c_{{ref}} \Delta_t+p_0 \Delta _z}\left(E_{0,j}^{n-1}+E_{1,j}^{n+1}\right)+\frac{2 p_0 \Delta_z}{c_{ref} \Delta_t+p_0 \Delta _z}\left(E_{0,j}^n+E_{1,j}^n\right) + \cr
		\fl	\frac{p_2 c_{ref}^2 \Delta_t^2 \Delta_z}{\Delta_y^2 \left(c_{ref} \Delta_t+p_0 \Delta _z\right)}\left\{\left(E_{0,j-1}^n+E_{0,j+1}^n-2E_{0,j}^n\right)+\left(E_{1,j-1}^n+E_{1,j+1}^n-2 E_{1,j}^n\right)\right\}.}
\end{equation}

\subsection{FDTD formulation of the ABC at $z=z_{end}$}
At $z=z_{end}$ or at $i=(N_z+\frac{1}{2})$
\begin{equation}
	\frac{\partial^2 E(r,t)}{\partial z\, \partial t}+\frac{p_0}{c_{ref}}\frac{\partial^2 E(r,t)}{\partial t^2} -p_2 c_{ref}\frac{\partial^2 E(r,t)}{\partial y^2}=0.
\end{equation}
Similarly, the time stepping equation can be obtained for the boundary node at $N_z+1$. 

\begin{eqnarray}
	\fl E_{N_z+1,j}^{n+1}=-E_{N_z,j}^{n-1}+\frac{c_{{ref}} \Delta_t-p_0 \Delta_z}{c_{{ref}} \Delta_t+p_0 \Delta _z}\left(E_{N_z+1,j}^{n-1}+E_{N_z,j}^{n+1}\right)+\frac{2 p_0 \Delta_z}{c_{{ref}} \Delta_t+p_0 \Delta _z}\left(E_{N_z+1,j}^n+E_{N_z,j}^n\right) \nonumber\\ 
	\fl+ \frac{p_2 c_{{ref}}^2 \Delta_t^2 \Delta_z}{\Delta_y^2 \left(c_{{ref}} \Delta_t+p_0 \Delta_z\right)} \left\{\left(E_{N_z+1,j-1}^n+E_{N_z+1,j+1}^n-2 E_{N_z+1,j}^n\right)+\right.\nonumber\\
	\fl \left. \left(E_{N_z,j-1}^n+E_{N_z,j+1}^n-2E_{N_z,j}^n\right)\right\}.
\end{eqnarray}

The third kind of boundary condition is the Total-field/Scattered-Field (TF/SF) boundary condition as shown in \fref{fig:gridsystemABC}(b). Such a boundary condition differentiates the computational domain into two regions, namely the Total field  (TF)  region and the scattered field (SF) region. The TF region contains the total field which is the sum of the incident wave and the scattered wave from the disorder. On the other hand, the $SF$ region contains only the scattered wave from the disorder as the source wave is subtracted from the total wave in these region. This is for the convenience of separating out and visualizing only the reflected wave from the disorder.

\section{Continuity equation and the estimation of total transmission and reflection without the STM}
\label{Appen:TotalT&R}
The local form of the continuity equation for any conserved quantity $q$ is given as
\begin{equation}
	\frac{\partial \rho (r,t) }{\partial t}+\nabla \cdot J(r,t)=0.
\end{equation}
If $q$ is the wave energy ($Joules$), then $\rho$ is the wave energy volume density  $(Joules/m^3)$ and $\vec{J}(r,t)$ is the wave current density ($Joules/(m^2 s)$). The scalar wave equation can then be casted in the form of the continuity equation as
\begin{equation}
	\eqalign{\fl	\frac{\partial }{\partial t} \underbrace{\left( \frac{1}{2 c^2}\left[\left(\frac{\partial E(r,t)}{\partial t}\right)^2\right]  + \frac{1}{2}|\nabla E(r,t)|^2 \right)}_{\rho(r,t)} + \nabla \cdot \underbrace{\left(  -\frac{\partial E(r,t)}{\partial t}\nabla E(r,t)\right)}_{\vec{J}(r,t)}=0.}
\end{equation}

In $2D$, $\vec{J}(r,t)=J_z(r,t)\hat{e}_z + J_y(r,t)\hat{e}_y$. Consider the case shown in \fref{fig:gridsystemABC}(b) where a dipole source excites a disordered slab without any STM modulations.  The total transmission $T$ and reflection $R$ of the slab disorder are obtained as 
\begin{eqnarray}
	T&=\frac{ \frac{1}{t_{cycle}}\int_{t=t_0}^{t=t_0+t_{cycle}}  \int_{y=0}^{y=W} J^{trans}_{z}(z,y,t)|_{(z,y)\in \Gamma_{R}} dy dt }{ \frac{1}{t_{cycle}}\int_{t=t_0}^{t=t_0+t_{cycle}} \int_{y=0}^{y=W} J^{source}_{z}(z,y,t)|_{(z,y)\in \Gamma_{S}} dy dt},\nonumber\\
	R&=\frac{ \frac{-1}{t_{cycle}}\int_{t=t_0}^{t=t_0+t_{cycle}}  \int_{y=0}^{y=W} J^{refl}_{z}(z,y,t)|_{(z,y)\in \Gamma_{L}} dy dt }{ \frac{1}{t_{cycle}}\int_{t=t_0}^{t=t_0+t_{cycle}} \int_{y=0}^{y=W} J^{source}_{z}(z,y,t)|_{(z,y)\in \Gamma_{S}} dy dt},
\end{eqnarray}
where $t_{cycle}=2\pi/\omega_{0}$ is the time period of the wave, $\Gamma_{S}$ is the incident slab surface, $J^{trans}_{z}(z,y,t)$ is the $z$ component of the wave current density associated with the transmitted wave in presence of scatterers, $J^{refl}_{z}(z,t)$ is the $z$ component of the wave current density associated with the reflected wave in presence of scatterers and $J^{source}_{z}(r,t)$ is $z$ component of the wave current density associated with the source wave, when no scattering perturbation against a background reference medium is present in the computational domain. Estimation of the source wave and $J^{source}_{z}(r,t)$ is done as a separate parallel run along with the FDTD time stepping in presence of the disorder.

\section{Phase sensitive lock-in detection of various STM diffraction orders}
\label{sec:PSDmethod}
The phase sensitive detection scheme presented in \cite{nihei2007frequency} involves a multi-frequency scenario where there is a constant frequency separation $\omega_{mod}$ between various harmonics components. Let the multi-frequency transmitted wave be $E_{trans}(r,t)|_{\Gamma_{R}}$ evaluated at the transmitting boundary $\Gamma_{R}$. Next, the amplitude and phase of the $\textbf{n}^{th}$ modulated wavefront with temporal frequency $\omega_\textbf{n}=\omega_{0} + \textbf{n}\omega_{mod}$ has to be estimated.  In order to do that, let there be two reference signals defined oscillating at the $\textbf{n}^{th}$ order temporal frequency such that,
\begin{eqnarray}
	\label{eq:inphase}
	& E^{ref_{0^\circ}}_\textbf{n}=E_{ref} cos(\omega_\textbf{n} t + \theta_{ref}),\\
	\label{eq:quadphase}
	&E^{ref_{90^\circ}}_\textbf{n}=E_{ref} cos(\omega_\textbf{n} t + \theta_{ref} + \pi/2),
\end{eqnarray}
where $\omega_\textbf{n}=\omega_{0} + \textbf{n}\omega_{mod}$, $E^{ref_{0^\circ}}_\textbf{n}$ is the in-phase reference signal at $\omega_\textbf{n}$ and $E^{ref_{90^\circ}}_\textbf{n}$ is the quadrature phase reference signal at $\omega_\textbf{n}$ corresponding to the $\textbf{n}^{th}$ harmonic frequency \cite{nihei2007frequency}. In the implementation of the coding, the amplitude $E_{ref}$ can be set to unity and the phase $\theta_{ref}$ to be zero, for convenience. Next step is to measure the spatially resolved temporal cross-correlation \cite{nihei2007frequency}  of the transmitted wave $E_{trans}(z,y,t)|_{(z,y)\in\Gamma_{R}}$, with that of the in-phase reference signal and the quadrature phase reference signal, over a time cycle where the time cycle is the beating time period $t_{beat}={2\pi}/{\omega_{mod}}$.  

\begin{eqnarray}
	\label{correqn:1}
	&\fl X^{corr}_\textbf{n}(z,y)|_{(z,y)\in\Gamma_R}=\frac{1}{t_{beat}}\int_{t=t_0}^{t=t_0 + t_{beat}} E^{ref_{0^\circ}}_\textbf{n}(t) E_{trans}(z,y,t)|_{(z,y)\in\Gamma_{R}} dt, \\
	\label{correqn:2}
	&\fl Y^{corr}_\textbf{n}(z,y)|_{(z,y)\in\Gamma_R}=\frac{1}{t_{beat}}\int_{t=t_0}^{t=t_0 + t_{beat}} E^{ref_{90^\circ}}_\textbf{n}(t) E_{trans}(z,y,t)|_{(z,y)\in\Gamma_{R}} dt.
\end{eqnarray}

Then, the amplitude $E^{trans}_\textbf{n}(z,y)|_{(z,y)\in\Gamma_{R}}$ and phase $\phi^{trans}_\textbf{n}(z,y)|_{(z,y)\in\Gamma_{R}}$ of the $\textbf{n}^{th}$ temporal component contained within the transmitted wave $E_{trans}(z,y,t)|_{(z,y)\in\Gamma_{R}}$ are estimated \cite{nihei2007frequency} as
\begin{eqnarray}
	\label{eq:lockinamp}
	E^{trans}_\textbf{n}(z,y)|_{(z,y)\in\Gamma_{R}}&=2 \sqrt{X^{corr}_\textbf{n}(z,y)^2 + Y^{corr}_\textbf{n}(z,y)^2}/E_{ref},\\
	\label{eq:lockinphase}
	\phi^{trans}_\textbf{n}(z,y)|_{(z,y)\in\Gamma_{R}} &=tan^{-1} (Y^{corr}_\textbf{n}(z,y)/X^{corr}_\textbf{n}(z,y)) + \theta_{ref}.
\end{eqnarray}

\section*{References} 
\bibliographystyle{unsrt}
\bibliography{bibliography}

\end{document}